\documentclass[aps,twocolumn, amsmath,amssymb,pra,superscriptaddress]{revtex4-2}

\usepackage{graphicx}
\usepackage{dcolumn}
\usepackage[dvipsnames]{xcolor}
\usepackage{bm}
\usepackage{setspace}
\usepackage{adjustbox}
\usepackage[normalem]{ulem}

\usepackage{empheq,etoolbox}

\usepackage[utf8]{inputenc}
\usepackage[english]{babel}
\usepackage[section]{placeins}
\usepackage{comment}
\graphicspath{ {./images/} }
\usepackage{indentfirst}
\usepackage{float}
\usepackage[a4paper, total ={6in,8in}]{geometry}

\DeclareMathOperator\Erf{Erf}
\DeclareMathOperator\Erfc{Erfc}

\usepackage{ulem}
\usepackage{color}

\newcommand{\delred}[1]{{\color{red}{\ifmmode\text{\sout{\ensuremath{#1}}}\else\sout{#1}\fi}}}

\begin{document}

\title{Dynamical galactic effects induced by stable vortex structure in bosonic dark matter
}

\author{K. Korshynska}
\affiliation{Department of Physics, Taras Shevchenko National University of Kyiv, 
64/13, Volodymyrska Street, Kyiv 01601, Ukraine}
\affiliation{Physikalisch-Technische Bundesanstalt (PTB), Bundesallee 100, D-38116 Braunschweig, Germany}
\author{Y.M. Bidasyuk}
\affiliation{Physikalisch-Technische Bundesanstalt (PTB), Bundesallee 100, D-38116 Braunschweig, Germany}
\author{E.V. Gorbar}
\affiliation{Department of Physics, Taras Shevchenko National University of Kyiv, 
64/13, Volodymyrska Street, Kyiv 01601, Ukraine}
\affiliation{Bogolyubov Institute for Theoretical Physics, 14-b Metrolohichna Street, Kyiv 03143, Ukraine}
\author{Junji Jia}
\affiliation{School of Physics and Technology, Wuhan University, 299 Bayi Roadd, Wuhan, Hubei Prov., China 430072}
\author{A.I. Yakimenko}
\affiliation{Department of Physics, Taras Shevchenko National University of Kyiv,
64/13, Volodymyrska Street, Kyiv 01601, Ukraine}
\affiliation{Dipartimento di Fisica e Astronomia ’Galileo Galilei’,
Universit`a di Padova, via Marzolo 8, 35131 Padova, Italy}
\affiliation{Istituto Nazionale di Fisica Nucleare, Sezione di Padova, via Marzolo 8, 35131 Padova, Italy}

\begin{abstract}
The nature of dark matter (DM) remains one of the unsolved mysteries of modern physics. An intriguing possibility is to assume that DM consists of ultralight bosonic particles in the Bose-Einstein condensate (BEC) state. We study stationary DM structures by using the system of the Gross-Pitaevskii and Poisson equations, including the effective temperature effect with parameters chosen to describe the Milky Way galaxy. We have investigated DM structure with BEC core and isothermal envelope. We compare the spherically symmetric and vortex core states, which allows us to analyze the impact of the core vorticity on the halo density, velocity distribution, and, therefore, its gravitational field. Gravitational field calculation is done in the gravitoelectromagnetism approach to include the impact of the core rotation, which induces a gravimagnetic field. As result, the halo with a vortex core is characterized by smaller orbital velocity in the galactic disk region in comparison with the non-rotating halo. It is found that the core vorticity produces gravimagnetic perturbation of celestial body dynamics, which can modify the circular trajectories. 
\end{abstract}

\maketitle

\section{Introduction}
The nature of DM particles remains one of the most fascinating puzzles of modern physics. The DM large-scale properties consistent with astrophysical observations are successfully explained by the cold dark matter model (CDM), which describes DM as a collisionless sufficiently cold perfect fluid. However, at smaller scales, the CDM encounters the cusp-core, missing satellites, and too-big-to-fail problems. One possibility to solve them is to assume that DM particles are ultra-light bosons as it is assumed in ultra-light dark matter (ULDM) models \cite{Ferreira}. Generically, these models are characterized by 
the suppression of the small-scale structures, the presence of cores, and dynamic effects which arise from the BEC formed in the central regions of galaxies. Such DM halo proposals were investigated in \cite{B_hmer_2007, https://doi.org/10.48550/arxiv.1611.09610,PhysRevD.86.064011, Hui_2017, rindler2014complex}. 

The ULDM model is supported indirectly by observations. For example, in cosmological simulations \cite{Schive_2014} it was found that the bosonic DM can indeed reproduce the observed distribution of matter at very large scales \cite{Matos_2000, PhysRevD.62.103517}, though the mass of such bosons should be extremely small. There have been also studies on some tensions of the ULDM with observational data from the rotation curves of galaxies including the Milky Way, which could probe the particle mass in the range $m = 10^{-22}- 10^{-21} $ eV \cite{bar2018galactic, de2020dynamical}. Furthermore, the viability of the ULDM model was studied with the stellar kinematics measurements in dwarf galaxies \cite{goldstein2022viability}. Another important piece of evidence is the DM nongravitational self-interaction, which has been recently reported for collisions of the clusters \cite{Harvey_2015, 2008arXiv0805.3827L}. In addition, the DM halo model must ensure the stability of a predicted halo. The stability of compact astrophysical objects which may be formed due to the Bose-Einstein condensation of ULDM was shown numerically \cite{2015PhRvD..91d4041M}. 

In the present paper, we discuss DM, which consists of ultra-light bosons with repulsive self-interaction. Such models make use of two macroscopic quantum phenomena: Bose-Einstein condensation and superfluidity. Bose-Einstein condensate in 
 the mean-field approximation is described by the Gross-Pitaevskii equation. By adding dissipation in the Gross-Pitaevskii equation one obtains a more general model, which includes the effective temperature effect and predicts that the ULDM halo consists of a BEC core and an isothermal envelope \cite{Chavanis-model}. Such core-envelope structure in the ULDM model was also discussed in \cite{launhardt2002nuclear, schonrich2015kinematic, portail2016dynamical, Schive_2014-core-halo}. Another important property, superfluidity, allows the quantization of the circulation and thus the possibility of the formation of vortices in the core of the halo. The central object of our study, the vortex, has a vanishing wavefunction at the vortex line, with a quantized circular flow around the vortex line 
\cite{Ferreira}. According to the recent numerical studies  \cite{Nikolaieva, Dmitriev} only the non-rotating soliton and single-charged vortex are stable, even being strongly perturbed. In the present work, we consider a DM halo, which consists of two regions - core and isothermal envelope, while the core could be either a soliton or a single-charged vortex.

Most of our knowledge about DM is based on its gravitational interaction with baryonic matter objects. Thus, testing the validity of the UDM theory requires a detailed investigation of the DM gravitational field. The DM density distribution, predicted by ULDM models, has been extensively studied in numerical simulations and applied in studies aimed at reconstructing the gravitational potential of DM halos for the Milky Way \cite{klypin2002lambdacdm} and dwarf galaxies \cite{walker2013dark}. In general, one can determine the gravitational field of the ULDM by solving the Einstein equations with the DM density and rotation flow as sources of the gravitational field, where rotation flow is induced by the BEC superfluidity. Thus, in the ULDM model, we should be able to deduce the impact of the superfluid DM rotation on the observations. The dominant effect of the vortex existence is due to the different core density distributions. Moreover, rotation flows produce $v/c$ and higher order effects, which can be taken into account in the gravitoelectromagnetism approach discussed in \cite{Toth,Hehl,Medina,Mashhoon,Wald} and used in our calculations below. The gravitoelectromagnetic formulation of a slowly rotating, self-gravitating, and dilute BEC intended for astrophysical applications in the context of DM halos was discussed in \cite{sarkar2018gravitationally}. As a rule, the gravimagnetic force is quite weak and does not affect significantly the dynamics of astrophysical systems.  However, in the central region of the BEC core, the DM density vanishes while the vortex flow velocity dramatically increases, which can affect the dynamics of luminous matter in the central region of the galaxies.


 In the present work, we calculate the DM gravitational field, which is needed for analysis of the observable predictions of the DM model, namely, to study how DM affects the movement of luminous matter. In our study, DM is the only source of a gravitational field, while luminous matter moves along geodesics, induced by DM. A more precise description of galactic kinematics is given by modeling the baryonic contribution to the gravitational potential which can distort the BEC soliton structures \cite{hayashi2020non, alexander2019dark}. Such a contribution was found to be significant for the Milky Way (MW) but not essential for the SPARC LSB galaxies \cite{bar2019ultralight}. In this paper, we will limit ourselves to some simple consequences of the ULDM model on the galactic kinematics, namely, rotation curves and deviation of circular trajectory, induced by the gravimagnetic force. The more detailed study in this direction is beyond the scope of the current paper, though it is an interesting perspective on further work.
 
The paper is organized as follows. In Sec.\ref{sec:model}, we develop the key parameters of our model, define the equations for halo structure, and formulate the gravitoelectromagnetism ansatz. In Sec.\ref{sec:halo}, we discuss the halo density profile for two stable core configurations and define the corresponding hydrodynamical velocity. In Sec.\ref{sec:gravielectric}, the gravielectric (Newtonian) field of the halo is calculated and the rotational curves are obtained. Sec.\ref{sec:gravimagnetic} provides gravimagnetic field calculations and our estimates of the gravimagnetic effect on circular trajectory. The results are summarized in Sec.\ref{sec:conclusions}.

\section{Model}
\label{sec:model}

\subsection{Ultra-light dark matter model and halo structure}

In this section, we briefly discuss the model, suggested in \cite{Chavanis-model}. The structure of the DM halo is described by the Gross-Pitaevskii-Poisson (GPP) equations, which define the dynamical evolution of self-gravitating BEC field $\psi$

\begin{eqnarray}
  i\hbar \frac{\partial \psi}{\partial t} &=& - \frac{\hbar^{2}}{2m}\Delta \psi + m\Phi_{\mathrm{g}}\psi + \frac{K \gamma m}{\gamma - 1}|\psi|^{2(\gamma - 1)} \psi  \nonumber\\
&& + \frac{m}{2}\left(\frac{3}{4\pi \eta_{\mathrm{0}}}\right)^{2/3}|\psi|^{4/3}\psi  + 2k_{\mathrm{B}}T \ln|\psi| \psi  \nonumber\\
&& - i \frac{\hbar}{2}\xi\left[\ln \left(\frac{\psi}{\psi^{*}}\right) - \left\langle \ln \left(\frac{\psi}{\psi^{*}}\right)\right\rangle \right]\psi,
\label{GP-equation}
\end{eqnarray}
\begin{equation}
  \Delta \Phi_{\mathrm{g}} = 4\pi G|\psi|^{2},
  \label{Poisson-equation}
\end{equation}
where $\langle X \rangle = \frac{1}{M} \int |\psi|^{2} X d\mathbf{r}$ is the spatial average over halo, $m$ is the bosonic particle mass, $\hbar$ denotes reduced Planck constant, $k_{\mathrm{B}}$ is Boltzmann constant. The first equation can be obtained by incorporating dissipative effects into the Schr{\"o}dinger equation by means of the theory of scale relativity. This generalization of the Schr{\"o}dinger equation means basically taking into account the interaction of the system with the external environment.  The model Eqs. (\ref{GP-equation}),(\ref{Poisson-equation}) were derived in \cite{Chavanis-relativity}, and we follow this approach in our current work. 

We consider the  BEC model with 
parameters $\gamma = 2$ and $K = \frac{2\pi a_{\mathrm{s}}\hbar^{2}}{m^{3}}$, where $a_{\mathrm{s}}$ denotes the $s$-wave scattering length of the self-interaction. The parameter $\eta_{\mathrm{0}}$ determines the equation of state of DM \cite{Poisson}. 
The first term on the right-hand side of Eq.(\ref{GP-equation}) is the kinetic term, and the second describes the interaction with the condensate gravitational
potential $\Phi_{\mathrm{g}}$. The third term takes into account the bosonic self-interaction (we will consider only the case $\gamma = 2$ which corresponds to binary collisions).  The fourth term accounts for the core, and the 
fifth term describes an isothermal envelope with effective temperature $T$ which surrounds the core. These terms can be derived from the Lynden-Bell theory of violent relaxation \cite{Chavanis-model}. The last term with $\xi < 0$ is a damping term 
and ensures that the system relaxes towards the equilibrium state.

An important feature of the Gross-Pitaevskii (GP) equation is that it satisfies the H-theorem, i.e., the free energy $F$ of the system decreases
\begin{equation*}
\dot{F} = -\xi \int \rho \mathbf{u}^{2} d\mathbf{r} \leq 0.
\end{equation*}
where $\rho = |\psi|^{2}$ denotes BEC density and $\mathbf{u} = \nabla S(\mathbf{r}, t)/m$ is the velocity field. These quantities are obtained by application of Madelung transformation, according to the expression $\psi(\mathbf{r}, t) = \sqrt{\rho(\mathbf{r}, t)} e^{iS(\mathbf{r},t)/\hbar}$, where $S(\mathbf{r},t)$ is the action. The negative sign of $\xi$ implies that the system relaxes towards the state with zero hydrodynamical velocity $\mathbf{u} = 0$. Therefore, a stationary vortex solution with nonzero $\mathbf{u}$ can be found only if we set
$\xi = 0$.

The free energy $F = E - TS_{\mathrm{B}}$ is expressed through the total energy $E$, the effective temperature $T$, and the Boltzmann entropy $S_{\mathrm{B}} = - k_{\mathrm{B}}\int (\rho/m)(\ln\rho - 1)d\mathbf{r}$. The total energy consists of 
the classical kinetic energy $\Theta_{\mathrm{c}} = 1/2\int \rho \mathbf{u}^{2}d\mathbf{r}$, the quantum kinetic energy $\Theta_{\mathrm{Q}} = 1/m\int \rho Q d\mathbf{r}$, the gravitational potential energy
$W = 1/2\int \rho \Phi_{\mathrm{g}} d\mathbf{r}$, and the internal energy of the self-interaction $U = K\int \rho^2 d\mathbf{r}$, 
$E_{\mathrm{0}} = \Theta_{\mathrm{c}} + \Theta_{\mathrm{Q}} + W + U$. Here $Q = -\frac{\hbar^{2}}{2m}\frac{\Delta \sqrt{\rho}}{\sqrt{\rho}}$ is the quantum potential. A stable equilibrium state corresponds to the minimum of the free energy $F$ at fixed total mass  $M$ of BEC. This gives the following condition of quantum hydrostatic equilibrium 
\cite{Chavanis-model}:
\begin{equation*}
\frac{\rho}{m}\nabla Q + \nabla P + \rho \nabla\Phi_{\mathrm{g}} + \frac{\rho}{2}\nabla\mathbf{u}^{2}  = 0,
\end{equation*}
where $P = K\rho^{2} + \rho \frac{k_{B}T}{m}$ is pressure due to the self-interaction and effective temperature.
Taking into account the Poisson equation (\ref{Poisson-equation}) and neglecting the quantum pressure term $Q$, we obtain the following equation of state
\begin{equation}
- 2K\Delta\rho - \frac{k_{\mathrm{B}}T}{m}\Delta \ln \rho = 4\pi G\rho + \frac{1}{2}\mathbf{u}^{2},
\label{isothermal equation}
\end{equation}
where $G$ is the gravitational constant. The solution of this equation is discussed in Sec. \ref{sec:halo}.

\subsection{Gravitoelectromagnetic approach}
\label{subsec:gravielectromagnetism}

To determine the gravitational field of DM halo we employ the well-known  gravitoelectromagnetism (GEM) approach \cite{Wald} which was previously applied to galactic structures in \cite{Toth,Medina,Mashhoon}. According to the GEM formalism, in the case of a test particle (which is luminous matter in our case) moving much slower than the speed of light $c$, it is convenient to represent the spacetime metric in the form
\begin{multline}
  dS^{2} = g_{\mu \nu}dx^{\mu}dx^{\nu}
  = \left(1 - \frac{2\Phi_{\mathrm{g}}}{c^{2}} \right)(dx^{0})^{2}  \\+\frac{4}{c^{2}} \left(\mathbf{A}_{\mathrm{g}}\mathbf{dx} \right)dx^{0} + \left(-1 - \frac{2\Phi_{\mathrm{g}}}{c^{2}} \right)\delta_{ij}dx^{i}dx^{j},
  \label{metrics}
\end{multline}
where $\Phi_{\mathrm{g}}$ and $\mathbf{A}_{\mathrm{g}}$ are the GEM scalar (gravielectric) and vector (gravimagnetic) potentials. For the gravitoelectromagnetic fields $\mathbf{E}_{\mathrm{g}}$ and $\mathbf{B}_{\mathrm{g}}$
\begin{equation}
   \mathbf{E}_{\mathrm{g}} = -\nabla{\Phi} -\frac{1}{2c}\partial_{t}\mathbf{A}_{\mathrm{g}},
   \label{gravielectric field}
\end{equation}
\begin{equation}
   \mathbf{B}_{\mathrm{g}} = \nabla \times \mathbf{A}_{\mathrm{g}},
   \label{gravimagnetic field}
\end{equation}
the Einstein equations imply the following relations:
\begin{equation*}
   \nabla \mathbf{E}_{\mathrm{g}} = 4\pi G \rho,
\end{equation*}
\begin{equation*}
   \nabla \times \mathbf{B}_{\mathrm{g}} = \frac{2}{c} \partial_{t} \mathbf{E}_{\mathrm{g}} + \frac{8\pi G}{c}\mathbf{j}.
\end{equation*}

Here sources of the gravitational field are mass density $\rho$ and matter current $\mathbf{j} = \rho \mathbf{u}$ ($\mathbf{u}$ is the matter velocity).

Since these equations are clearly analogous to those in the electromagnetic theory, their solutions have a form similar to  Maxwell's theory
\begin{equation}
     \Phi_{\mathrm{g}}(\mathbf{r}) = G\int_{\Omega}\frac{\rho(\mathbf{r'})  d^{3}r'}{|\mathbf{r} - \mathbf{r'}|},
     \label{gravielectric potential}
\end{equation}
\begin{equation}
     \mathbf{A_{\mathrm{g}}}(\mathbf{r}) = \frac{2G}{c}\int_{\Omega}\frac{\rho(\mathbf{r'}) \mathbf{u}(\mathbf{r'}) d^{3}r'}{|\mathbf{r} - \mathbf{r'}|},
     \label{gravimagnetic potential}
\end{equation} 
where integration proceeds over $\mathbf{r'}$ occupied by DM particles, $\rho(\mathbf{r'})$ is the condensate density and $\mathbf{u}(\mathbf{r'})$ is the BEC velocity at $\mathbf{r'}$. $\mathbf{r}$ are coordinates associated with the test particle, moving along geodesics in the BEC gravitational field.

Finally, the geodesic movement for a test particle, which corresponds to the spacetime metric in the GEM form, 
 \begin{equation*}
   \frac{d^{2}x^{i}}{dt^{2}}  = \frac{\partial \Phi_{\mathrm{g}}}{\partial x_{i}}+ \frac{2}{c}\frac{dA_{\mathrm{g}}^{i}}{dt} - \frac{2}{c}\left(\frac{\partial \mathbf{A}_{\mathrm{g}}}{\partial x_{i}}\frac{d\mathbf{x}}{dt}\right) 
\end{equation*}
can be equivalently described as the classical motion $m\ddot{\mathbf{x}}=\mathbf{F}_{\mathrm{g}}$ in the gravitoelectromagnetic analog of the Lorentz force
\begin{equation}
   \mathbf{F}_{\mathrm{g}} = -m \left(\mathbf{E}_{\mathrm{g}} + \frac{2}{c}\mathbf{v}\times\mathbf{B}_{\mathrm{g}} \right) = m (\mathbf{a}_{\mathrm{E}} + \mathbf{a}_{\mathrm{B}}),
   \label{Lorenz force}
\end{equation}
where $\mathbf{v}$ is the particle velocity and $m$ is its mass. Here we introduced gravielectric $\mathbf{a}_{\mathrm{E}} = - \mathbf{E}_{\mathrm{g}}$ and gravimagnetic $\mathbf{a}_{\mathrm{B}} = -\frac{2}{c}\mathbf{v}\times\mathbf{B}_{\mathrm{g}} $ components of acceleration.

\section{Halo density profile}
\label{sec:halo}

The model, based on the generalized GPP equations (see Eqs. (\ref{GP-equation}), (\ref{Poisson-equation})) describes the core-envelope structure of DM halo with a dense core and diffuse isothermal envelope. The model yields the following equation of state for the ULDM $P = K\rho^{2} + \rho\frac{k_{B}T}{m}$ (see Sec. \ref{sec:model}). Thus, one can conclude, that in the core region equation of state is approximately $P = K\rho^{2}$, because the weak self-interaction dominates over effective temperature impact due to large density. That is why the latter will be neglected in the discussion of the core states. On the contrary, in the isothermal envelope region, we have the equation of state $P = \rho\frac{k_{B}T}{m}$, which means that the effective temperature term plays a crucial role there.

Based on these considerations, we calculate the halo density in two steps. Firstly, we reproduce the numerical result for the total density of the non-rotating halo (see the original result in \cite{Chavanis-model}), which defines density distribution in the isothermal envelope region. This step is needed as a starting point to define isothermal envelope density distribution and to compare the discussed in \cite{Chavanis-model} $s=0$ solitonic core with the new case of vortex core $s=1$. Secondly, under the assumption that core and envelope do not interact, we discuss the core density profile separately by means of variational ansatz \cite{Chavanis}. This way we will study the spherically symmetric ($s=0$) and the single-charged vortex ($s=1$) solutions for the core density distribution. 

\subsection{Isothermal envelope}

In the first case of a non-rotating core, we can set $\mathbf{u} = 0$ , and then the Eq. (\ref{isothermal equation}) simplifies
\begin{equation*}
   - \frac{4\pi a_{\mathrm{s}}\hbar^{2}}{m^{3}}\Delta\rho - \frac{k_{\mathrm{B}}T}{m}\Delta \ln \rho = 4\pi G\rho,
\end{equation*}
where we took into account that $K = \frac{4\pi a_{\mathrm{s}}\hbar^{2}}{m^{3}}$.

It is convenient to introduce the density function and the radial coordinate $\rho = \rho_{\mathrm{c}}e^{-f}$, $y = r/r_{\mathrm{0}}$, where
\begin{equation}
r_{\mathrm{0}} = \sqrt{\frac{k_{\mathrm{B}}T}{4\pi G \rho_{\mathrm{c}} m}}
\label{r0scale}
\end{equation}
and $\rho_{\mathrm{c}}$ defines the density at the center. The equation of state can be rewritten in the following form:
\begin{equation}
\frac{d^{2}f}{dy^{2}} + \frac{2}{y}\frac{df}{dy} = \frac{\chi \left(\frac{df}{dy}\right)^{2} + 1}{\chi + e^{f}},
\label{state-equation}
\end{equation}
where $\chi = 4\pi a_{\mathrm{s}} \hbar^{2}\rho_{\mathrm{c}}/(m^{2}k_{\mathrm{B}}T)$. The boundary conditions are $f(0) = 0$ and $\frac{df}{dy}(0) = 0$, which define the boundary conditions for the density function $\rho(0) = \rho_{\mathrm{c}}$ and
$\frac{d\rho(0)}{dr} = 0$. 
We solve Eq. (\ref{state-equation}) numerically for different values of $\chi$ and present solutions in Fig. \ref{fig full density} (a). 
The isothermal envelope density distribution is defined as $\rho =  \rho_{0}e^{-f} = \rho_{0}f_{\mathrm{N}}(r)$, where $f$ is a numerical solution of Eq. (\ref{state-equation}).

The profile has a solitonic core and an isothermal envelope whose density decreases as $\rho(r) \sim k_{\mathrm{B}}T/(2\pi Gmr^{2}) = v_{\infty}^{2}/(4\pi Gr^{2})$ \cite{Chavanis-model} in agreement with observations (here $v_{\infty}$ is the constant rotational velocity in the large distance limit). The existence of a BEC core in the ULDM model was also discussed in \cite{launhardt2002nuclear, schonrich2015kinematic, portail2016dynamical, Schive_2014-core-halo}. 

The possible physical origin of the core-envelope structure could be the merger of two-state configurations when the total system tends to a virialized state, and the obtained averaged profile has a core and a tail structure \cite{Guzm_n_2018}. The process of halo formation usually undergoes gravitational cooling \cite{guzman2006gravitational}, which is discussed in \cite{seidel1994formation, guzman2006gravitational}. Gravitational cooling process for inreitially quite arbitrary density profiles leads to relaxation and virialization through the emission of scalar field particles \cite{guzman2016behavior}. The resulting profile has the same dense core and diffuse envelope structure.

In the case $s=1$, the hydrodynamical velocity $\mathbf{u}$ does not vanish in the inner region due to the existence of the vortex. The definition of the velocity profile in the isothermal halo region is a complicated task. One would expect that there is an intermediate region between the core and isothermal envelope, where the hydrodynamical velocity is small but nonzero, and at large enough distances, we should have $\mathbf{u} = 0$. This is due to the divergent mass of the isothermal envelope, which therefore cannot rotate in order for kinetic energy to be finite. For an estimate, we simply put $\mathbf{u} = 0$ in the whole isothermal envelope region. This approximation can be justified by the negligibly small density of the isothermal envelope in comparison with the core density, so its rotation would have no sufficient impact on the system. Hence the density profile in the envelope region remains unchanged. {Thus, to define isothermal envelope density distribution we use the numerical solution for $\rho = \rho_{0}f_{\mathrm{N}}(r)$, obtained earlier in the case of non-rotating core.} The density profile in the core region will be discussed in the next section in detail.

To reproduce the Milky Way halo mass $M= 1.3 \times 10^{12}M_{\odot}$ and radius $R_{\mathrm{halo}} = 287 kpc$ \cite{Posti}, taking into account the model described in \cite{Chavanis-model}, we choose the following values of the
particle mass $m = 2.92 \times 10^{-22} eV/c^{2} = 0.52 \times 10^{-57} kg$, scattering length $a_{\mathrm{s}} = 8.17 \times 10^{-77} m$, effective DM temperature $T = 5.09 \times 10^{-25} K$, central density in the spherical
case $\rho_{\mathrm{c}} = 0.34 \times 10^{-17} \frac{kg}{m^{3}}$ and distance scaling parameter $r_{\mathrm{0}} = 0.071 kpc$. Then $\chi = 20$ and the temperature of BEC of such ultralight bosons is much larger than the effective temperature.  For the spherically symmetric case, this yields the core with mass $M_{\mathrm{c}} = 6.39 \times 10^{10}M_{\odot}$ and radius $R_{\mathrm{c}} = 1 kpc$.

\subsection{Core stationary states}
The dynamics of self-gravitating BEC of $N$ weakly interacting bosons with mass $m$ is described by the GPP system of equations with the term, corresponding to the effective temperature impact:
\begin{multline} 
  i\hbar\frac{\partial\psi}{\partial t} = \left(-\frac{\hbar^{2}}{2m}\nabla^{2} + gN|\psi|^{2} + m\Phi_{g} \right. \\ \left.+ 2k_{B}T \ln\Big|\frac{\psi}{\psi_{0}}\Big|\right)\psi
  \label{GPP 1}
\end{multline}
\begin{equation}
  \nabla^{2}\Phi_{g} = 4\pi GmN|\psi|^{2}
  \label{GPP 2}
\end{equation}
where $g = \frac{4 \pi \hbar^{2}a_{s}}{m}$ is the coupling strength that corresponds to the two-particle interaction, $a_{s}$ is the s-wave scattering length, $\Phi$ is the gravitational potential and G is gravitational constant.

The GPP system of Eqs.(\ref{GPP 1}) and (\ref{GPP 2}) includes three crucial physical parameters: particle mass $m$, the total number of particles $N$ (or, equivalently, total mass $M$) and coupling strength $g$ (or, equivalently, self-interaction constant $\frac{\lambda}{8\pi} = \frac{a_{\mathrm{s}}}{\lambda_{\mathrm{c}}}$, where $\lambda_{\mathrm{c}} = \frac{\hbar}{mc}$ is the Compton wavelength of bosons) \cite{Chavanis}. 

The GPP system of equations is invariant under the transformation $t = \lambda_{*}^{2}t'$, $\mathbf{r} = \lambda_{*}\mathbf{r}'$, $\psi = \lambda_{*}^{-2}\psi'$, $\Phi_{\mathrm{g}} = \lambda_{*}^{-2}\Phi'_{\mathrm{g}}$, $g = \lambda_{*}^{2}g'$, where $\lambda_{*} > 0$, which allows us to scale-out the coupling constant to $g=1$.

In order to simplify calculations, it is convenient to introduce dimensionless variables and wave function
\begin{equation} 
  i\frac{\partial\psi}{\partial t} = \left(-\frac{1}{2}\nabla^{2} + |\psi|^{2} + \Phi_{\mathrm{g}} + T_{\mathrm{eff}}\ln|\psi|\right)\psi,
  \label{GP equation dimensionless}
\end{equation}
\begin{equation}
  \nabla^{2}\Phi_{\mathrm{g}} = |\psi|^{2},
  \label{Poisson equation dimensionless}
\end{equation}
where the dimensional variables are related to the dimensionless ones as follows: $\mathbf{r} = \mathbf{r}_{\mathrm{ph}}{L}$, $t = \omega_{*} t_{\mathrm{ph}}$, $\Phi_{\mathrm{g}} = \left(\frac{\lambda_{\mathrm{c}}}{L}\right)^{2}\frac{\Phi_{\mathrm{g ph}}}{c^{2}}$, and $\psi = \frac{\lambda}{8\pi}\left(\frac{m_{\mathrm{Pl}}}{m}\right)^{2}\sqrt{4\pi GM}\frac{\hbar}{mc^{2}}\psi_{\mathrm{ph}}$. Here the distance and time scaling parameters are $L = \lambda_{\mathrm{c}}\frac{m_{\mathrm{Pl}}}{m}\sqrt{\frac{\lambda}{8\pi}} = \frac{m_{\mathrm{Pl}}\hbar}{m^{2}c}\sqrt{\frac{\lambda}{8\pi}} = 0.99 \times 10^{19}m = 0.32 kpc$ and $\omega_{*} = \frac{c\lambda_{\mathrm{c}}}{L^{2}} = 2.08 \times 10^{-15} s^{-1}$. The dimensionless effective temperature parameter is $T_{\mathrm{eff}} = \frac{2k_{\mathrm{B}}T}{\omega_{*}\hbar}$ and will be neglected in the following discussion because the corresponding term $T_{\mathrm{eff}}\ln|\psi|$ is negligibly small in the core region. {Therefore, we neglect the temperature effects in the analysis of the BEC core density distribution .}

For the BEC core mass $M_{\mathrm{c}} = 6.39 \times 10^{10}M_{\odot}$ and radius $R_{\mathrm{c}} = 1 kpc$, we solve the GPP equations (\ref{GP equation dimensionless}) and (\ref{Poisson equation dimensionless}) by using the variational ansatz in cylindrical coordinates $r,z$
\begin{equation}
\psi(r, \phi, z) = A \left(\frac{r}{R}\right)^{s}e^{-\frac{r^{2}}{2R^{2}} - \frac{z^{2}}{2(R\eta)^{2}} + is\phi}.
\label{variational ansatz}
\end{equation}
 Here $R$ and $\eta$ are variational parameters, which will be fixed later. Constant $A$ is fixed by the normalization condition
\begin{equation}
A = \sqrt{\frac{N_{\mathrm{0}}}{\pi^{3/2}\eta R^{3}s!}},
\label{normalisation condition}
\end{equation}   
the cases $s=0,1$ are considered, and $N_{\mathrm{0}}$ is defined by the core mass
\begin{equation*}
    N_{\mathrm{0}} = 4\pi\frac{M_{\mathrm{c}}}{m_{\mathrm{Pl}}}\sqrt{\frac{\lambda}{8\pi}} = 2.55 \cdot 10^{4},
\end{equation*}
where $m_{\mathrm{Pl}} = \sqrt{\frac{\hbar c}{G}}$ is the Planck mass and $\lambda/(8\pi) = 1.21 \times 10^{-91}$ is the self-interaction coupling constant.

The dimensionless quantities and the physically observed ones are related as follows:
\begin{equation}
    R_{\mathrm{c}} = R_{\mathrm{99}}L = \frac{m_{\mathrm{Pl}}\hbar}{m^{2}c}\sqrt{\frac{\lambda}{8\pi}} R_{\mathrm{99}},
    \label{total radius}
\end{equation} 
\begin{multline}
\rho = M|\psi_{\mathrm{ph}}|^{2} =  \frac{M}{L^{3}N_{\mathrm{0}}}|\psi|^{2}  \\= \rho_{\mathrm{0}}\left(\frac{r}{R}\right)^{2s}e^{-\frac{r^{2}}{R^{2}} - \frac{z^{2}}{(R\eta)^{2}}},
\label{dimension density}
\end{multline} 
where $R_{\mathrm{99}}$ is the dimensionless radius which contains $99$ percent of the mass of the core \textcolor{black}{(the variational analysis gives  $R_{99} \approx 2.38R$ in the case of solitonic core and $R_{99} \approx 2.58R$ in the case of vortex core)}, $\rho$ is the condensate density, and $\rho_{\mathrm{0}} = MA^{2}/(L^{3}N_{\mathrm{0}})$ is the density scaling parameter. $R_{\mathrm{c}}$ denotes the total radius of the core in physical units.

Using the variational ansatz for the BEC wave function (\ref{variational ansatz}), we obtain the energy \cite{Chavanis}:
\scriptsize
\begin{multline}
E =   \int d^{3}\mathbf{r} \psi^{*}(\mathbf{r},t)\left(-\frac{1}{2}\nabla^{2} + |\psi|^{2} + \Phi_{\mathrm{g}}\right)\psi(\mathbf{r},t) \\=\epsilon \left(\frac{N_{\mathrm{0}}(1 + 2\eta^{2}(1+s))}{4R^{2}\eta^{2}} + \frac{N_{\mathrm{0}}^{2}\Gamma(s+1/2)}{4\sqrt{2}\pi^{2}R^{3}\eta\Gamma(s+1)}  \right. \\-\left.\frac{N_{\mathrm{0}}^{2}}{8\pi R} \int_{0}^{\infty}
\Erfc\left(\frac{k_{*}\eta}{\sqrt{2}}\right)L_{\mathrm{s}}^{2}\left(\frac{k_{*}^{2}}{4}\right) e^{-\frac{k_{*}^{2}(1 - \eta^{2})}{2}} dk_{*}\right).
\label{free-energy}
\end{multline} 
\normalsize
where $\Gamma(x)$ denotes the Gamma function, $\Erfc(x)$ is the complementary error function and $L_{\mathrm{s}}(x)$ is the Laguerre polynomials. Here $\epsilon = (\hbar^{2}/4\pi m_{\mathrm{Pl}} \lambda_{\mathrm{c}}^{2})(8\pi/\lambda)^{3/2}$ is characteristic energy, which does not depend on variational parameters.

In what follows, we will use $r_{\mathrm{0}} = 2.18 \times 10^{18} m = 0.071 kpc = 0.22 L$ as the distance scaling parameter.

In the subsection below, we investigate the case $s=0$.

\subsubsection{Non-rotating spherically-symmetric core}

In this case, the BEC wave function in Eq. (\ref{variational ansatz}) depends only on radial distance $r$ in spherical coordinates
\begin{equation}
\psi(r) = A e^{-\frac{r^{2}}{2R^{2}}}
\label{s=0 var ansatz}
\end{equation}
and the density function (see Eq. (\ref{dimension density})) equals
\begin{equation}
\rho(r) = \rho_{\mathrm{0}} e^{-\frac{r^{2}}{R^{2}}}.
\label{density-spherical}
\end{equation}

In what follows, $r$ will denote spherical distance, when the $s=0$ case is discussed.

We should relate $R$ and the BEC core radius $R_{\mathrm{c}}$ which is defined through $M_{\mathrm{c}}= \frac{4}{\pi}\rho_{\mathrm{0}}R_{\mathrm{c}}^{3}$ \cite{Chavanis}. Since
$\rho_{\mathrm{0}} = M_{\mathrm{c}}A^{2}/(L^{3}N_{\mathrm{0}})$, the numerical result for halo density (see Fig. \ref{fig full density} a) gives $\frac{R_{\mathrm{c}}}{LR} = 1.64$ or $R = 8.66$ in the $r_{\mathrm{0}}$ scale. It is interesting to compare the obtained $R$ with its value in the variational analysis method used in \cite{Chavanis}. Substituting $\eta =1$ and $s=0$ in the energy functional in Eq. (\ref{free-energy}), we get
\begin{multline*}
    \frac{E}{\epsilon} = \frac{3N_{\mathrm{0}}}{4R^{2}} + \frac{N_{\mathrm{0}}^{2}}{4\sqrt{2}\pi^{3/2}R^{3}} \\ -  \frac{N_{\mathrm{0}}^{2}}{8\pi R} \int_{0}^{\infty} \Erfc\left(\frac{k_{*}}{\sqrt{2}}\right)  dk_{*}. 
\end{multline*} 
Its extremum is defined by the equation
\begin{equation}
R^{2}-\frac{6\sqrt{2}\pi^{3/2}}{N_{\mathrm{0}}}R - 3 =0
\end{equation} 
that gives $R = 1.73$ or $R = 7.86$ in the $r_{\mathrm{0}}$ scale. Thus, $R_{\mathrm{c}} = 0.9\,kpc$ (see Eq. (\ref{total radius})) and, therefore, the variational analysis method and numerical calculation (see Fig. \ref{fig full density} (a)) are in a good agreement.

\subsubsection{Rotating axially-symmetric core}

In the case $s=1$ (see Eq.(\ref{variational ansatz})), we have a wave function, which depends on cylindrical coordinates $r, z, \phi$
\begin{equation}
  \psi(r, \phi, z) = A \frac{r}{R} e^{-\frac{r^{2}}{2R^{2}} - \frac{z^{2}}{2(R\eta)^{2}} + i\phi}
  \label{s=1 var ansatz}
\end{equation}
and the density function equals
\begin{equation}
  \rho(r,  z) = \rho_{\mathrm{0}} \frac{r}{R} e^{-\frac{r^{2}}{R^{2}} - \frac{z^{2}}{(R\eta)^{2}}},
  \label{s=1 density}
\end{equation}
where $A$ is given by Eq. (\ref{normalisation condition}).

\begin{figure*}[t!]
  \begin{center}
    \includegraphics[width=\textwidth]{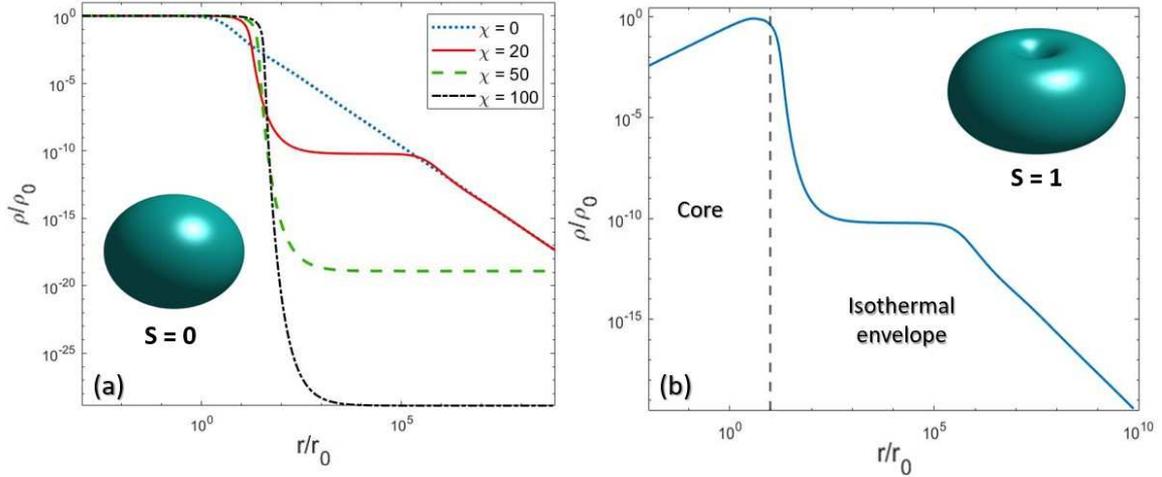}
    \caption{Halo density profile $\rho/\rho_{0}$ as a function of dimensionless $r/r_{0}$ coordinate in the plane $z=0$, both $x$ and $y$ axes have log scale.  The cyan insets in both plots show 3D density isosurfaces of the corresponding BEC cores. Left panel (a) shows the halo with the BEC core in a soliton state ($s=0$). Three curves correspond to different values of parameter  $\chi = 4\pi a_{\mathrm{s}} \hbar^{2}\rho_{\mathrm{c}}/(m^{2}k_{\mathrm{B}}T)$, so that while increasing $\chi$ one decreases effective temperature $T$ and vice versa. Right panel (b) shows the halo with the core in a vortex state ($s=1$, $\chi$ = 20). \textcolor{black}{Note, we investigate in detail the isothermal envelope for $\chi$ = 20, which is consistent with observations for the Milky Way}. The black dashed line divides the distribution into two parts: the inner region with a rotating core and the outer region composed of an isothermal envelope.}
     \label{fig full density}
    \end{center} 
\end{figure*}

The dimensionless total energy in Eq. (\ref{free-energy}) for $s=1$ reads
\begin{multline*}
    \frac{E}{\epsilon} = \frac{N_{\mathrm{0}}(1 + 4\eta^{2})}{4R^{2}\eta^{2}} + \frac{N_{\mathrm{0}}^{2}}{8\sqrt{2}\pi^{3/2}R^{3}\eta} \\ - \frac{N_{\mathrm{0}}^{2}}{8\pi R} \int_{0}^{\infty} \Erfc \left(\frac{k_{*}\eta}{\sqrt{2}}\right)
    \left(1 - \frac{k_{*}^{2}}{4}\right)^{2} e^{-\frac{k_{*}^{2}(1 - \eta^{2})}{2}} dk_{*}.
\end{multline*} 
Equations of an extremum of the total energy with respect to $\eta$ and $R$ yield the solution $\eta = 1.464,$ and $R = 1.226$ in the $L$ scale. In the $r_{\mathrm{0}}$ scale, we have $R = 5.57$.  To determine the core density distribution, we use the variational analysis result. We assume that the core interacts negligibly weakly with the isothermal envelope. Therefore, for the isothermal envelope region, we use the numerical distribution $f_{\mathrm{N}}(r_{\mathrm{sph}}) = f_{\mathrm{N}}(\sqrt{r^{2} + z^{2}})$ (see Fig. \ref{fig full density} (a)), derived under $\mathbf{u} =0 $ condition.

Thus, we obtain (see Fig. \ref{fig full density} (b))
\begin{equation}
 \rho(r,z) =  \rho_{0}  
\begin{cases}
 1.92 \frac{r}{R}e^{-\frac{r^{2}}{R^{2}} - \frac{z^{2}}{(R\eta)^{2}}}, \frac{r_{\mathrm{sph}}}{r_{\mathrm{0}}} \leq \frac{R_{\mathrm{c}}}{r_{\mathrm{0}}}\\    f_{\mathrm{N}}\left(\frac{r_{\mathrm{sph}}}{r_{\mathrm{0}}}\right), \frac{r_{\mathrm{sph}}}{r_{\mathrm{0}}} > \frac{R_{\mathrm{c}}}{r_{\mathrm{0}}} ,
\end{cases}
\label{s=1 full density-distribution}
\end{equation}
where $r = \sqrt{x^{2} + y^{2}}$ and $z$ are cylindrical coordinates and $r_{\mathrm{sph}} = \sqrt{x^{2} + y^{2} + z^{2}}$. Here $\rho_{\mathrm{0}}$ is the spherical halo central density. \textcolor{black}{The spherically symmetric isothermal envelope density $\rho(r,z) = \rho_{0} f_{\mathrm{N}}\left({r_{\mathrm{sph}}}/{r_{\mathrm{0}}}\right)$ is found numerically by solving Eq. (\ref{state-equation}). The total core radius $R_{\mathrm{c}}$ is defined by Eq. (\ref{total radius}).}

By using $\mathbf{u} = \frac{\mathbf{j}_{\mathrm{ph}}}{\rho}$ and the particle current
\begin{multline*}
    \mathbf{j}_{\mathrm{ph}} = -\frac{i\hbar}{2m}(\psi_{\mathrm{ph}}^{*} \mathbf{\nabla}\psi_{\mathrm{ph}} - \psi_{\mathrm{ph}} \mathbf{\nabla}\psi_{\mathrm{ph}}^{*}) \\= \frac{\hbar}{m}\frac{|\psi_{\mathrm{ph}}|^{2}}{r}\mathbf{e}_{\phi},
\end{multline*}
we find the velocity distribution $\mathbf{u}(\mathbf{r})$ of DM particles
\begin{equation}
    \mathbf{u} = \frac{\frac{\hbar}{m}\frac{|\psi_{\mathrm{ph}}|^{2}}{r}}{|\psi_{\mathrm{ph}}|^{2}} \mathbf{e}_{\phi} = \frac{\hbar}{m}\frac{1}{r}\mathbf{e}_{\phi} = \alpha \frac{cr_{\mathrm{0}}}{r}\mathbf{e}_{\phi},
\label{velocity-distribution}
\end{equation}
where $\alpha = \hbar/(mr_{\mathrm{0}}c) = 0.31 \cdot 10^{-3}$. Obviously, the velocity of condensate particles increases while approaching the center of the vortex. Note that there is an inner region where the velocity becomes of the 
order of $c$ and, therefore, this region cannot be described by making use of the gravitoelectromagnetism ansatz (see Appendix A for explanation). This region is limited by the radial distance $r = \alpha r_{\mathrm{0}} = 2.2 \times 10^{-5}\,kpc$.

In the two following sections, by using the formalism of GEM, we describe particle movement in the gravitational field of DM in the $s=0,1$ states aiming to understand how baryonic matter particles interact with the proposed DM .

\section{Gravielectric field and rotation curves}
\label{sec:gravielectric}

In this section, we obtain numerical results for the gravielectric (Newtonian) component of the DM halo gravitational field. Having calculated the field, we analyze the rotation curves, predicted by the model in the cases of soliton and vortex core.

To determine the gravielectric potential in the case of a non-rotating halo we use the numerically obtained density distribution (see Fig. \ref{fig full density} (a)). In the case of a rotating axially symmetric halo, the mass density distribution is shown in Fig. \ref{fig full density} (b).

\begin{figure}[!htb]
   \centering
   \includegraphics[width=\columnwidth]{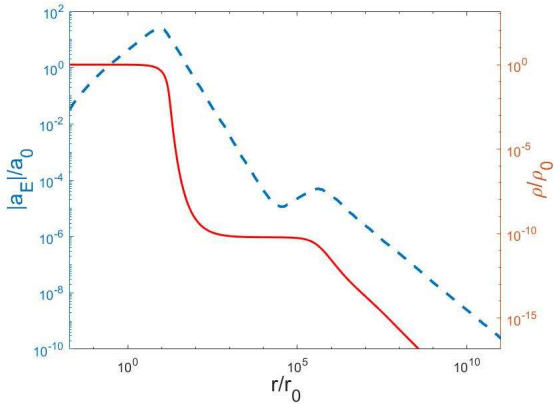}
    \caption{The radial component of gravielectric field $a_{\mathrm{E}}/a_{\mathrm{0}}$ (blue dashed line)  and density (red solid line) of the non-rotating halo ($s=0$ core) as functions of the dimensionless $r/r_{\mathrm{0}}$ coordinate, both $x$ and $y$ axes have log scale. Here $a_{\mathrm{0}}=5.38 \times 10^{-13} \frac{km}{s^{2}}$, $r_{0} = 71 pc$.}
    \label{fig s=0 gravielectric field}
\end{figure}

In the spherically symmetric case of non-rotating halo ($s=0$), only the radial component of the gravielectric field is not zero (see Eq.(\ref{gravielectric potential})) and the corresponding gravielectric acceleration $\mathbf{a}_{\mathrm{E}} = - \mathbf{E}_{\mathrm{g}} = a_{\mathrm{E}}\mathbf{e}_{r}$ (see Eq.(\ref{Lorenz force})) is presented in Fig. \ref{fig s=0 gravielectric field}. The acceleration at large distances behaves like $a_{\mathrm{E}}/a_{\mathrm{0}} = 82.66r_{\mathrm{0}}/r$, i.e., $a_{\mathrm{E}}= 9.3 \times 10^{-29} \frac{\,kpc^{2}}{s^{2}} \times 1/r$. Here $a_{\mathrm{0}} =G\rho_{\mathrm{0}} r_{\mathrm{0}}=  5.38 \times 10^{-13} km/s^{2}$.  In the core region, where the density distribution is described by the variational ansatz (\ref{density-spherical}), the gravielectric potential and the corresponding acceleration can be found analytically
\begin{equation*}
     \frac{1}{r^{2}}\frac{\partial}{\partial r} r^{2}\frac{\partial}{\partial r}  \Phi_{\mathrm{g}} = -4\pi G \rho_{\mathrm{0}} e^{-\frac{r^{2}}{R^{2}}}.
\end{equation*} 
The general solution is given by
\begin{equation*}
   \Phi(a) = - 4\pi G \rho_{\mathrm{0}}R^{2}\left(\frac{c_{\mathrm{1}}}{r} + c_{\mathrm{2}} - \frac{R\sqrt{\pi}\Erf(r/R)}{4r}\right).
\end{equation*} 
where $\Erf(x)$ denotes the error function and $c_{1}$, $c_{2}$ are constants.

\begin{figure}[!htb]
    \centering
    \includegraphics[width=\columnwidth]{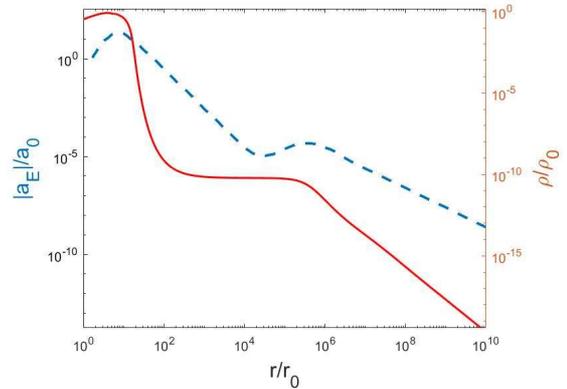}
    \caption{The radial component of gravielectric acceleration $a_{\mathrm{Er}}/a_{\mathrm{0}}$ (blue dashed line) and density (red line)  of the rotating halo ($s=1$ core) as functions of dimensionless $r/r_{\mathrm{0}}$ coordinate in the $z=0$ plane, both $x$ and $y$ axes have log scale. Here $a_{\mathrm{0}}=5.38 \times 10^{-13} \frac{km}{s^{2}}$, $r_{0} = 71 pc$.}
    \label{fig s=1 gravielectric radial z=0}
\end{figure}

\begin{figure*}[t!]
\begin{center}
\includegraphics[width=\textwidth]{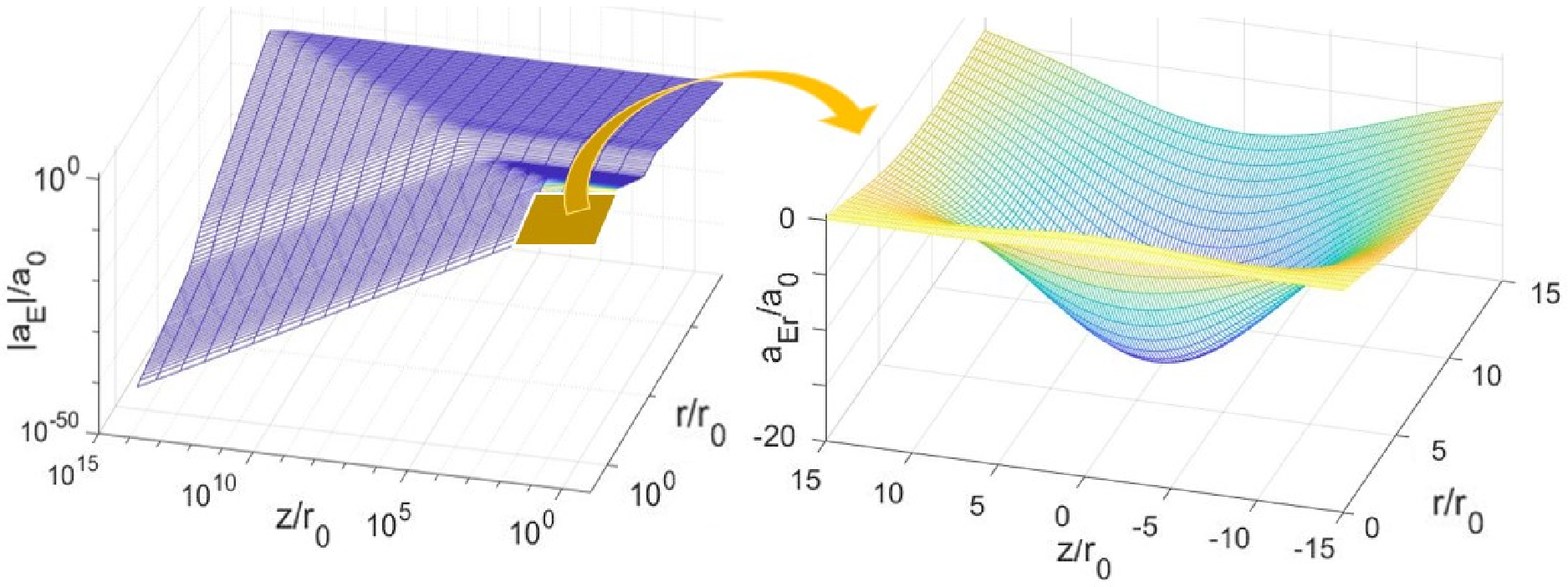}
\caption{The radial component of gravielectric acceleration $a_{\mathrm{Er}}/a_{\mathrm{0}}$ induced by the rotating halo ($s=1$ core) as a function of dimensionless $r/r_{\mathrm{0}}$ and $z/r_{\mathrm{0}}$ coordinates. Here $a_{\mathrm{0}}=5.38 \times 10^{-13} \frac{km}{s^{2}}$, $r_{0} = 71 pc$. The left panel shows the isothermal envelope region with the three axes in the log scale and the right panel is a zoom-in of the core region.}
\label{fig s=1 gravielectric radial}
\end{center}
\end{figure*}

\begin{figure*}[t!]
  \begin{center}
    \includegraphics[width=\textwidth]{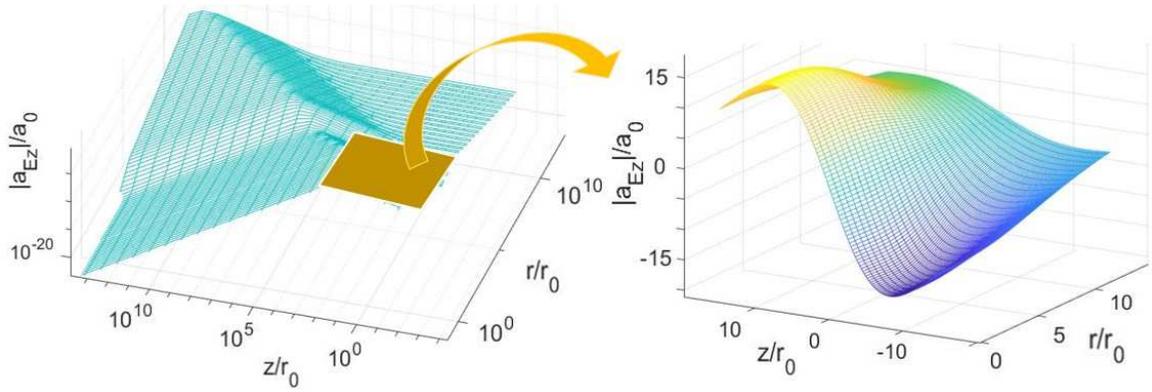}
    \caption{The $z$-component of gravielectric acceleration $a_{\mathrm{Ez}}/a_{\mathrm{0}}$ induced by the rotating halo ($s=1$ core) as a function of dimensionless $r/r_{\mathrm{0}}$ and $z/r_{\mathrm{0}}$ coordinates. Here $a_{\mathrm{0}}=5.38 \times 10^{-13} \frac{km}{s^{2}}$, $r_{0} = 71 pc$. The left panel shows the isothermal envelope region with the three axes in the log scale and the right panel is a zoom-in of the core region.}
    \label{fig s=1 gravielectric z}
    \end{center}
\end{figure*}

We can set $c_{\mathrm{2}} = 0$. At a large distance, the gravielectric potential of the halo must be equal to the potential of a body with the same mass $M = \pi^{3/2}\rho_{\mathrm{0}}R^{3}$. This implies that $c_{\mathrm{1}} = 0$. Thus, $\Phi(r)$ is 
completely determined and we have the radial acceleration
\begin{multline*}
   \mathbf{a}_{\mathrm{E}}(r) = \nabla\Phi_{\mathrm{g}}(r)  \\= \pi G \rho_{\mathrm{0}}R^{3}\left( \frac{2e^{-r^{2}/R^{2}}}{Rr}
   - \frac{\sqrt{\pi}\Erf\left({\frac{r}{R}}\right)}{r^{2}}\right)\mathbf{e}_{r}.
\end{multline*} 
Clearly, $a_{\mathrm{E}}$ has a maximum at $r = R =8.66$ in the $r_{\mathrm{0}}$ scale in agreement with the radial gravielectric acceleration shown in Fig. \ref{fig s=0 gravielectric field}.

Gravielectric field in the case of vortex core has radial and $z$ components in cylindrical coordinates, namely $\mathbf{a}_{\mathrm{E}} = a_{\mathrm{Er}}\mathbf{e}_{r} + a_{\mathrm{Ez}}\mathbf{e}_{z}$. They are illustrated in Figs. \ref{fig s=1 gravielectric radial} and \ref{fig s=1 gravielectric z}, respectively. The radial dependence of the gravielectric radial acceleration in the $z=0$ plane is shown in Fig. \ref{fig s=1 gravielectric radial z=0}. Notice that at $r \approx 0.81r_{\mathrm{0}} = 0.058\,kpc$ the acceleration projection changes sign, hence test particles are repelled in the interior region and attracted in the exterior region. This result stems from the geometry of the considered doughnut-shaped halo with a hole.

\begin{figure}[htp]
    \centering
   \includegraphics[width=\columnwidth]{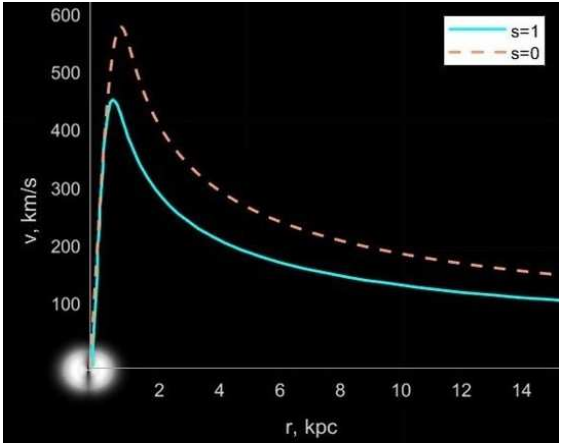}
    \caption{The  rotation (Kepler) velocity $v$ in the $z=0$ plane as a function of the radial distance $r$. The pink dashed line corresponds to the non-rotating spherical halo ($s=0$ core) and the cyan solid line to the rotating halo ($s=1$ core). The background represents a gradient plot of the density distribution in $s=1$ case}
    \label{fig rotational curves}
\end{figure}

Now we aim to determine the impact of the gravitational field of the DM halo on the movement of celestial bodies in the Milky Way galaxy. According to our model (see Sec. \ref{sec:halo}) density distribution depends on the state of the core, which must lead to a difference between the rotation curves, which they induce. To demonstrate how the gravielectric acceleration induces rotation in the $s=0$ and $s=1$ cases, we present the rotation velocity $v$ in the $z=0$ plane as a function of the radial distance $r$ in Fig. \ref{fig rotational curves}. The new result here is the curve in the case $s=1$, while $s=0$ case was discussed earlier in \cite{Chavanis-model}. The two halos with $s=0$ and $s=1$ core have equal mass, which is the observed mass of DM halo in the Milky Way, according to the model discussed in Sec. \ref{sec:halo}. The numerical results indeed show that at large distances the corresponding rotational curves have the same asymptotic. Note that, the gravielectric force in $s=1$ case changes its sign at $r = 0.81r_{\mathrm{0}} = 0.058 \,kpc$. Hence, at distances less than $0.058 \,kpc$ there are no stable rotation orbits in the rotating halo model. However, the stable orbits are possible if one includes not only DM but also the other sources of the gravitational field, namely, the baryonic galactic bulge and the supermassive black hole in the central region of the galaxy.

\section{Gravimagnetic field in the BEC core}
\label{sec:gravimagnetic}

In this section, we obtain numerical results for the gravimagnetic (first post-Newtonian) component of the DM halo gravitational field (see Subsec. \ref{subsec:gravielectromagnetism} of Sec.\ref{sec:model}). The component is induced by a moving source, hence, it is nonzero only in the second case of the DM halo with a vortex core.

\begin{figure}[!htb]
    \centering
    \includegraphics[width=\columnwidth]{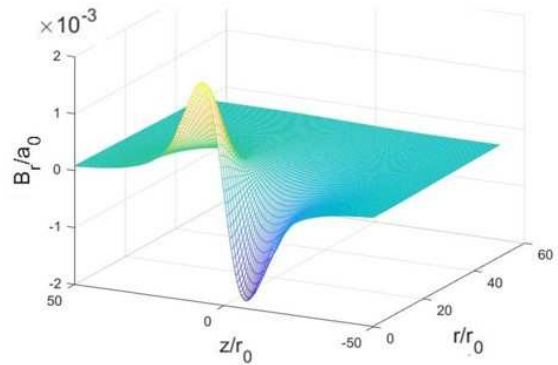}
    \caption{The radial component of gravimagnetic field $B_{r}/a_{\mathrm{0}}$ induced by the rotating core as a function of dimensionless $r/r_{\mathrm{0}}$ and $z/r_{\mathrm{0}}$ coordinates. Here $a_{\mathrm{0}}=5.38 \times 10^{-13} \frac{km}{s^{2}}$, $r_{0} = 71 pc$.}
    \label{fig s=1 gravimagnetic radial}
\end{figure}

\begin{figure}[h]
    \centering
    \includegraphics[width=\columnwidth]{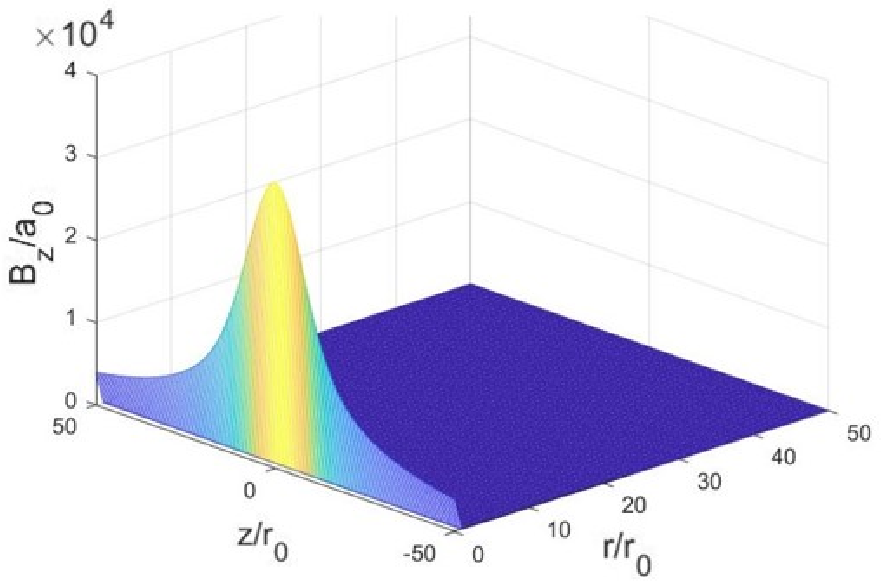}
    \caption{The $z$-component of gravimagnetic field $B_{z}/a_{\mathrm{0}}$ induced by the rotating core as a function of dimensionless $r/r_{\mathrm{0}}$ and $z/r_{\mathrm{0}}$ coordinates. Here $a_{\mathrm{0}}=5.38 \times 10^{-13} \frac{km}{s^{2}}$, $r_{0} = 71 pc$.}
    \label{fig s=1 gravimagnetic z}
\end{figure}

\begin{figure}[h]
    \centering
    \includegraphics[width=\columnwidth]{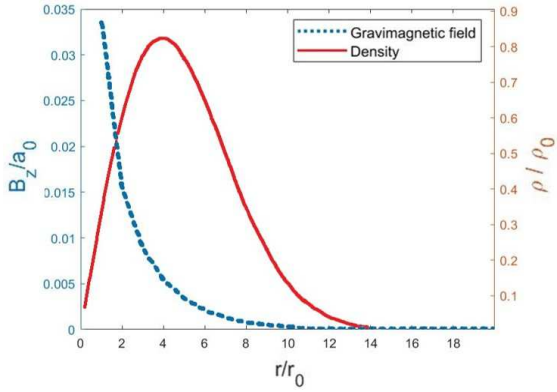}
    \caption{The $z$-component of the gravimagnetic field $B_{z}/a_{\mathrm{0}}$ (blue dashed line) and density (red line) of the rotating core as functions of dimensionless $r/r_{\mathrm{0}}$ in $z=0$ plane. Here $a_{\mathrm{0}}=5.38 \times 10^{-13} \frac{km}{s^{2}}$, $r_{0} = 71 pc$.}
    \label{fig s=1 gravimagnetic z at z=0}
\end{figure}

To determine the gravimagnetic potential in the case of a rotating axially symmetric halo we use the mass density and velocity distributions given by Eqs. (\ref{s=1 full density-distribution}) and
(\ref{velocity-distribution}). The calculation is based on Eqs. (\ref{gravimagnetic potential}) and (\ref{gravimagnetic field}). The results of numerical integration for radial and $z$-components of the gravimagnetic field, $\mathbf{B} = B_{r}\mathbf{e}_{r} + B_{z}\mathbf{e}_{z}$, are shown in Figs. \ref{fig s=1 gravimagnetic radial} and \ref{fig s=1 gravimagnetic z}, respectively. Fig. \ref{fig s=1 gravimagnetic z at z=0} displays the $z$-component of the gravimagnetic field $B_{z}$ in the $z=0$ plane (the radial component of the gravimagnetic field equals 
zero in this plane).

Having determined the gravimagnetic field, we can calculate the corresponding acceleration of the test particle. Using Eq.(\ref{Lorenz force}), we have
\begin{multline*}
   \mathbf{a}_{\mathrm{B}}(\mathbf{r} = (a,b,k)) = - \frac{2}{c}v\mathbf{e}_{\phi}\times\mathbf{B}_{\mathrm{g}}  \\
    =-1.38\alpha G\rho_{\mathrm{0}}r_{\mathrm{0}}\frac{v}{c}(B_{r}(a,b,k)\mathbf{e}_{r} + B_{z}(a,b,k)\mathbf{e}_{z})  \\ = a_{\mathrm{Br}}(a,b,k)\mathbf{e}_{r} + a_{\mathrm{Bz}}(a,b,k)\mathbf{e}_{z}.
\end{multline*}
where $a = r/r_{0}$, $b = \phi$, $c = z/r_{0}$ are rescaled cylindrical coordinates.

This allows us to estimate the impact of the gravimagnetic field on stars' motion. In the case of the Milky Way galaxy, $v = v_{\mathrm{0}} + \gamma_{\mathrm{0}}r_{\mathrm{0}}a$ if $a<a_{\mathrm{break}}$ and $v = v_{1} + \gamma_{1}r_{\mathrm{0}}a$ for $a \geq a_{\mathrm{break}}$ \cite{Obreja}. Constants $\gamma_{0}$, $\gamma_{1}$, $a_{\mathrm{break}}$, and $v_{1}$ are different 
for the thick and thin galactic disks' velocity profiles. Setting $R_{\mathrm{break}} = r_{\mathrm{0}} a_{\mathrm{break}} =5\,kpc$ and $v_{\mathrm{0}}$=0 in both cases gives the values of parameters presented in Table I. This approximation is valid up to
$13\,kpc = 180 r_{\mathrm{0}}$ \cite{Obreja}. We should emphasize here that $v$ includes only the component of velocity directed along $\mathbf{e}_{\phi}$ and does not include the component along $\mathbf{e}_{r}$. It is important to 
distinguish the $\phi$-component and the absolute value of the whole velocity when dealing with sufficiently non-circular elliptic orbits.

\begin{table*}
\caption{\label{tab:table3}Parameters of the Milky Way’s rotational velocity profiles \cite{Obreja}.}
\begin{ruledtabular}
\begin{tabular}{cccc}
 Galactic disk& $v_{1}$ $[km s^{-1}]$& $\gamma_{0}  [km s^{-1} kpc^{-1}]$&$\gamma_{1} [km s^{-1} kpc^{-1}]$\\ \hline
 thin disk& 236.71 & 45.41 & -1.93 \\
 thick disk & 206.93 & 39.086 & -2.30
\end{tabular}
\end{ruledtabular}
\end{table*}

\begin{figure}[ht]
    \centering
   \includegraphics[width=\columnwidth]{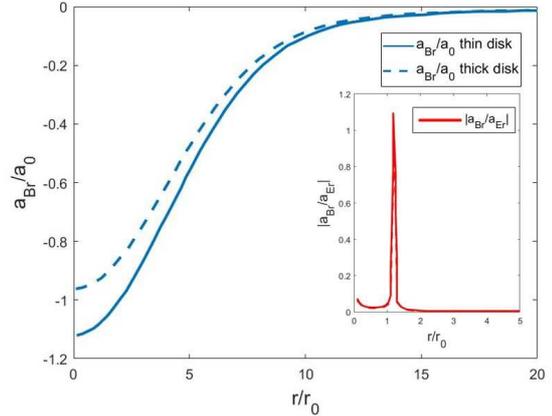}
    \caption{The radial component of gravimagnetic acceleration (solid and dashed blue lines) $a_{\mathrm{Br}}/a_{\mathrm{0}}$ for thin and thick disks, respectively, and the absolute value of the ratio of gravimagnetic acceleration to gravielectric $|a_{\mathrm{Br}}/a_{\mathrm{Er}}|$ (both for thin and for thick disks) as a function of $r/r_{\mathrm{0}}$. Here $a_{\mathrm{0}}=5.38 \times 10^{-13} \frac{km}{s^{2}}$, $r_{0} = 71 pc$.}
    \label{fig gravimagnetic and gravielectric 1}
\end{figure}

According to Eq.(\ref{Lorenz force}), the gravimagnetic acceleration in galactic plane $c=0$ can be estimated as
\begin{equation*}
   \mathbf{a}_{\mathrm{B}} =-1.38\alpha G\rho_{0}r_{\mathrm{0}}\frac{v_{i} + \gamma_{i}r_{0}a}{c}B_{r}(a,b,0)\mathbf{e}_{r},
\end{equation*}
where $i=0$ for $a<a_{\mathrm{break}}$ and $i=1$ for $a \geq a_{\mathrm{break}}$. The corresponding plot is shown in Fig. \ref{fig gravimagnetic and gravielectric 1}. The spike on the red curve, which shows the modulus of the ratio of the gravimagnetic acceleration to the 
gravielectric one $|a_{\mathrm{Br}}/a_{\mathrm{Er}}|$ appears because the gravielectric acceleration changes sign at $r = 0.81r_{0} = 0.058\,kpc$.

\begin{figure}[ht]
    \centering
    \includegraphics[scale=0.85]{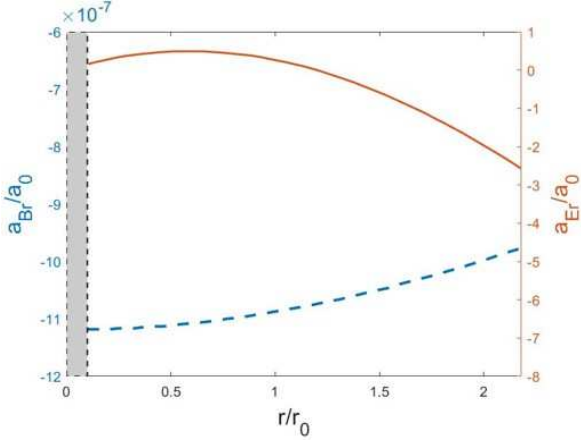}
    \caption{The radial component of gravimagnetic acceleration (dashed blue line) $a_{\mathrm{Br}}/a_{\mathrm{0}}$  and gravielectric acceleration (red line) $a_{\mathrm{Er}}/a_{\mathrm{0}}$ in the inner region of halo. Here the grey region corresponds to $r<0.1r_{\mathrm{0}}$, where the gravimagnetic approximation is not valid. Here $a_{\mathrm{0}}=5.38 \times 10^{-13} \frac{km}{s^{2}}$, $r_{0} = 71 pc$.}
    \label{fig graviamgnetic versus gravielectric}
\end{figure}

It is interesting that $a_{\mathrm{Br}}(a)$ tends to a constant in the $a \ll 1$ limit (see Fig. \ref{fig graviamgnetic versus gravielectric}). This directly follows from the analytical expression. In the interior region $r < 5 \,kpc$, we have
\begin{equation*}
  \frac{a_{\mathrm{Br}}}{a_{\mathrm{0}}} =-1.38\alpha \frac{ \gamma_{0}r_{\mathrm{0}}a}{c}B_{r}(a, b, 0).
\end{equation*}
In the $a\ll 1$ limit, we find

\begin{multline*}
B_{r}(a, b, 0)  \\ \approx
 \frac{2\pi}{a}\int_{0}^{\infty}dx\int_{-\infty}^{\infty}dz \frac{x^{2}}{\sqrt{x^{2} + z^{2}}}e^{-\frac{x^2}{R^{2}} - \frac{z^{2}}{(R\eta)^{2}}}.
\end{multline*}

The last integral can be calculated numerically which yields
\begin{multline*}
  \frac{a_{\mathrm{Br}}}{a_{\mathrm{0}}}  \approx -1.38\alpha \frac{ \gamma_{0}r_{0}}{c}\times 247 = 9.8 \times10^{-7}.
\end{multline*}

We see that in the case under consideration the gravimagnetic acceleration indeed tends to be a constant in the $a \ll 1$ limit.

The gravimagnetic field calculations performed in this section allow us to obtain some testable predictions of the model. According to numerical results for $\mathbf{B}_{\mathrm{g}}$ and $\mathbf{E}_{\mathrm{g}}$, the gravielectric force changes its sign at $r = 0.81r_{0} = 0.058 \,kpc$, and the gravimagnetic force component is attractive or repulsive, depending on the direction of 
the motion. The acceleration in the polar coordinates $(r, \phi)$ is given by $\mathbf{a}=(\ddot{r} - r\dot{\phi}^{2})\mathbf{e}_{r} + (r\ddot{\phi} + 2\dot{r}\dot{\phi})\mathbf{e}_{\phi} $. Then the equations of motion for a star take the form
\begin{equation}
    \frac{d^{2}r}{dt^{2}} = r\left(\frac{d\phi}{dt}\right)^{2} - E_{r} - \frac{2B_{z}r}{c}\frac{d\phi}{dt},
    \label{Kepler with gravimagnetic radial}
\end{equation}
\begin{equation}
    r\frac{d^{2}\phi}{dt^{2}} = \frac{2B_{z}}{c}\frac{dr}{dt} -  2\frac{dr}{dt}\frac{d\phi}{dt}.
    \label{Kepler with gravimagnetic angular}
\end{equation}

Since the gravielectric acceleration dominates over the gravimagnetic one, it suffices to take the latter into account as a perturbation. Therefore, we treat $B_{\mathrm{g}}$ as the first-order perturbation and expand $\phi(t)$ 
and $r(t)$ around the solution $r_{\mathrm{c}}$ and $\phi_{\mathrm{c}}$ determined by the gravielectric acceleration. For $r = r_{\mathrm{c}} + \delta r$ and $\phi = \phi_{\mathrm{c}} + \delta \phi$, in the zeroth-order, we have the 
Kepler problem equations with $E_{r}(r_{\mathrm{c}})$ calculated numerically in Sec.\ref{sec:gravielectric}. The corresponding solutions are elliptic orbits. For simplicity, we will consider only the case of circular orbits 
$r_{\mathrm{c}}(\phi) = r_{\mathrm{c}} = const$. By substituting $E_{r}(r_{\mathrm{c}} + \delta r) \approx E(r_{\mathrm{c}}) + \frac{dE}{dr}(r_{\mathrm{c}}) \delta r$ in Eqs. (\ref{Kepler with gravimagnetic radial}) and (\ref{Kepler with gravimagnetic angular}), we obtain
\begin{equation*}
    \frac{d^{2}\delta r}{dt^{2}} = w_{0}^{2}\delta r + 2r_{\mathrm{c}}w_{0}\frac{d\delta \phi}{dt} - \frac{dE_{r}}{dr}\Big|_{r_{\mathrm{c}}} \delta r - \frac{2B_{z}}{c}r_{\mathrm{c}}w_{0}, 
\end{equation*}
\begin{equation*}
    r_{\mathrm{c}}\frac{d^{2}\delta \phi}{dt^{2}} =  -2w_{0}\frac{d\delta r}{dt}.
\end{equation*}
where $w_{0} = \frac{d \phi_{\mathrm{c}}}{dt}$ is angular frequency, induced by gravielectric field. Thus, it can be explicitly written as $w_{0}^{2} = \frac{E_{r}(r_{\mathrm{c}})}{r_{\mathrm{c}}}$
Integrating the second equation, we get
\begin{equation*}
    \frac{d\delta \phi}{dt} =  -\frac{2w_{0}}{r_{\mathrm{c}}}\delta r,
\end{equation*}
where we set the integration constant to zero. Substituting this relation in the first equation, we find
\begin{equation*}
   \frac{d^{2}\delta r}{dt^{2}} = -\left( 3w_{0}^{2}  + \frac{dE_{r}}{dr}\Big|_{r_{\mathrm{c}}} \right) \delta r - \frac{2B_{z}}{c}r_{\mathrm{c}}w_{0}.
\end{equation*}
From the numerical result, we see that $f(E)$ is positive and tends to zero at a large distance. Then, for
$3\frac{E_{r}(r_{\mathrm{c}})}{r_{\mathrm{c}}} + \frac{dE_{r}}{dr}\Big|_{r_{\mathrm{c}}} = \Omega^{2} > 0$, we find solutions
\begin{equation*}
   \delta r = -\frac{2B_{z}r_{\mathrm{c}}w_{0}}{c\Omega^{2}} +J\sin(\Omega (t - t_{0})),
\end{equation*}
\begin{equation*}
   \delta \phi = \delta \phi_{\mathrm{c}} +\frac{4B_{z}w_{0}^{2}}{c\Omega^{2}}t + \frac{2w_{0}J}{\Omega r_{\mathrm{c}}}\cos(\Omega( t - t_{0})),
\end{equation*}
where $\delta r_{\mathrm{c}}$, $J$, and $t_{0}$ are defined by the corresponding initial conditions.

It is interesting to estimate $\Omega^2$ at some distance $r_{\mathrm{c}}$, e.g., $r_{\mathrm{c}} = 8\,kpc = 113r_{\mathrm{0}}$ which is the distance of the Sun from the center of the galaxy. Then we have
$\Omega = \sqrt{0.0017 \times a_{0}/r_{0}} = 6.48 \times 10^{-16} s^{-1}$ (the corresponding period is $T = 3.1 \times 10^{8}\,y$),  $B_{z} = 2.76  \times 10^{-8}a_{0}$, and $ \delta r = - 2B_{z}r_{\mathrm{c}}w_{0}/c\Omega^{2} = -2.7 \times 10^{-8} r_{0} = - 0.38 a.u.$. The latter distance is approximately equal to
80 solar radii. The angular frequency is shifted by the value $4B_{z}w_{0}^{2}/c\Omega^{2} = 4.8 \times 10^{-25} s^{-1}$ (the corresponding period is $T = 4.2 \times 10^{17} y$).

\section{Conclusions}
\label{sec:conclusions}

We investigated the model of DM halo with BEC core composed of ultra-light bosonic particles. Solving the generalized GPP equations for self-gravitating BEC we obtained the density profile of the DM halo and analyzed its core and envelope structure. The density and velocity profiles were found for two types of  stable structures with topological charges ($s=0$ and $s=1$) of the BEC core. 

Using this DM halo description, we investigated its gravitational field and the impact of this field on the baryonic matter.
The key result of our paper is that the observable effects, predicted by the ULDM halo model, depend on the state of the core. In particular, solitonic and vortex cores yield different density and velocity distributions and thus different gravitational fields. The doughnut-like density distribution (vanishing at the vortex core) and vortex flows (rapidly increasing at the vortex axis) of the BEC core can significantly modify both gravielectric and gravimagnetic components of the gravitational field. We described the gravitational fields of these two core configurations by using the gravimagnetism approach. A dominant component of the gravitational field is the gravielectric (Newtonian) one, which generates the rotation of celestial bodies in the galaxy. The rotational velocity  induced by the halo with vortex is smaller close to the core region but has the same asymptotics at large distances in comparison with the non-rotating halo. 

The first post-Newtonian component of the gravitational field, which is called gravimagnetic, is induced by the rotation of the BEC vortex core and appears only in the model of a rotating halo. Although, as expected, the gravimagnetic acceleration is much weaker than the gravielectric one, it can affect the dynamics of baryonic matter in the halo, especially in its inner region. In our simplified perturbation approach for circular orbit gravimagnetic field yields radius and frequency shift, and can also induce trajectory oscillations, depending on initial conditions.

There are several possible directions in which the present study could be extended. An analysis of gravitational fields beyond the gravimagnetic approach is required in the central region of the galaxy, due to the high rotational velocity of BEC there. Furthermore, according to astrophysical observations, there is a supermassive black hole in the center of our galaxy whose presence should be taken into account. Finally, the gravitational effects of baryonic matter should be included in further studies.

\section{Acknowledgments}

The authors are grateful to Yelyzaveta Nikolaieva, Sebastian Ulbricht, Stanislav Vilchinskii,  and Luca Salasnich for useful discussions and comments.  A.Y. acknowledge support from BIRD Project "Ultracold atoms in curved geometries" of the University of Padova.
 
\section*{Appendix A}

Let us discuss the self-consistency of our model, which makes use of the GEM approach to describe the first post-Newtonian contribution to the gravitational field potential. We assumed that a test particle (celestial body acted upon by the gravitational field) propagates with a non-relativistic speed $v$ so that all terms of higher than linear order in $O(v/c)$ can be neglected in the equations of motion. As to DM, we describe it by using the nonlinear Schr\"{o}dinger equation with gravitational potential $\Phi_{\mathrm{g}}$.

Since the hydrodynamical velocity in the vortex (the state with $s=1$) is $u(r) = \alpha {cr_{0}}/{r}$, it increases at small $r$ and attains at $r \sim \alpha r_{0}$ values of the order of $c$. Obviously, the Newtonian treatment is not applicable in this region. Therefore, we use the Klein-Gordon equation in order to describe the relativistic equation of motion of bosons, as follows:
\begin{equation} 
  \nabla_{\alpha}\nabla^{\alpha} \phi + \left[\left(\frac{mc}{\hbar}\right)^{2}  - U(|\phi|^{2}) \right]\phi = 0,
\end{equation}
where $U = \frac{2m}{\hbar^{2}}gN|\phi|^{2}$ and $\phi$ is the scalar field. We neglect the effective temperature because only the core region is investigated (the hydrodynamical velocity $u(r)$ is nonzero only in the core region) and $\nabla_{\alpha}$ denotes covariant derivative in curved space-time.

The metric in the GEM approach reads (here all notations are the same as in Subsec. \ref{subsec:gravielectromagnetism})
\begin{multline*}
  dS^{2} = g_{\mu \nu}dx^{\mu}dx^{\nu} = \left(1 - \frac{2\Phi_{g}}{c^{2}} \right)(dx^{0})^{2} \\+ \frac{4}{c^{2}} \left(\mathbf{A}_{g}\mathbf{dx} \right)dx^{0} + \left(-1 - \frac{2\Phi_{g}}{c^{2}} \right)\delta_{ij}dx^{i}dx^{j}
\end{multline*}
and the Laplace operator is given by
\begin{equation*}
\nabla_{\alpha}\nabla^{\alpha}\phi = \frac{1}{\sqrt{-g}}\partial_{\alpha}(\sqrt{-g} g^{\alpha \beta} \partial_{\beta}\phi)
\end{equation*}
where $g = det(g_{\mu \nu}) \approx -1$.
Then we have
\begin{multline*}
\nabla_{\alpha}\nabla^{\alpha}\phi = \frac{1}{c^{2}}\left(1 - \frac{2\Phi_{g}}{c^{2}}\right)\partial^{2}_{t}\phi  - \frac{2A^{i}_{g}}{c^{3}}\partial_{t}\partial_{i}\phi \\- \frac{2}{c^{3}} \partial_{i}(A^{i}_{g} \partial_{t}\phi) - \partial_{i}\left[ \left(1 + \frac{2\Phi_{g}}{c^{2}}\right)\delta^{ij} \partial_{j}\phi\right],
\end{multline*}
where fields $\Phi_{g}$ and $\mathbf{A}_{g}$ are time-independent. Taking into account the gauge condition $\partial_{i}A^{i}_{g} = 0$, we find
\begin{multline*}
\nabla_{\alpha}\nabla^{\alpha}\phi = \frac{1}{c^{2}}\left(1 - \frac{2\Phi_{g}}{c^{2}}\right)\partial^{2}_{t}\phi \\ - \frac{4A^{i}_{g}}{c^{3}}\partial_{t}\partial^{i}\phi + \partial_{i}\left[ \left(1 + \frac{2\Phi_{g}}{c^{2}}\right) \partial_{i}\phi\right].
\end{multline*}

To obtain a nonrelativistic approximation of the Klein-Gordon equation we represent the scalar field in the form $\phi = e^{imc^{2}t/\hbar}\psi$. Substituting this expression in the Klein-Gordon equation and multiplying by $e^{-imc^{2}t/\hbar}$ we get
\begin{multline*} 
  \frac{1}{c^{2}}\left(1 - \frac{2\Phi_{g}}{c^{2}}\right)\left[\partial_{t}^{2}\psi + \frac{2imc^{2}}{\hbar}\partial_{t}\psi - \left(\frac{mc^{2}}{\hbar}\right)^{2}\psi\right] \\- \frac{4A^{i}_{g}}{c^{3}}\left[\partial_{t}\partial^{i}\psi + \frac{imc^{2}}{\hbar}\partial^{j}\psi \right] +\left[1 + \frac{2\Phi_{g}}{c^{2}}\right]\partial^{j}\partial_{j}\psi \\+ \frac{2}{c^{2}}\partial^{j}\Phi_{g} \partial_{j}\psi + \left[\left(\frac{mc}{\hbar}\right)^{2}  - \frac{2m}{\hbar^{2}}U(|\psi|^{2}) \right]\psi = 0.
\end{multline*}
Neglecting terms of order of $(u/c)^{2}$ and higher ($A_{g} \sim u/c$), we obtain
\begin{multline*} 
   \frac{2im}{\hbar}\partial_{t}\psi - \left(\frac{mc}{\hbar}\right)^{2}\psi  + 2\Phi_{g} \left(\frac{m}{\hbar}\right)^{2}\psi + \partial^{j}\partial_{j}\psi  \\+ \left[\left(\frac{mc}{\hbar}\right)^{2}  - \frac{2m}{\hbar^{2}}U(|\psi|^{2}) \right]\psi = 0
\end{multline*}
Finally, after some straightforward simplifications, we derive the Schr{\"o}dinger equation in the form
\begin{equation*} 
  i\hbar \partial_{t}\psi  = \left( -  \frac{\hbar^{2}}{2m} \partial^{j}\partial_{j} +   m \Phi_{g}   +  U(|\psi|^{2})\right)\psi.
\end{equation*}

Thus, we conclude that the model is self-consistent if we take into account only terms up to $u/c$, or, equivalently, in the region, where the hydrodynamical velocity of vortex is not relativistic ($u \ll c$).

\bibliography{Bibl}

\end{document}